\documentclass[aps,pra,twocolumn,groupedaddress,showpacs,superscriptaddress]{revtex4-1}

\newcommand{\hide}[1]{}

\usepackage{amsmath,amssymb,graphicx}
\usepackage{stmaryrd} 
\usepackage{pdfpages} 
\usepackage{float} 
\usepackage{url}
\usepackage{cmap}  
\usepackage{lineno} 
\usepackage{yhmath}  
\usepackage{ulem} 
\usepackage{subfigure}
\usepackage{mathcomp} 
\usepackage{nicefrac}
\usepackage[nice]{units}

\newcommand{\Lyh}{\nicefrac{L_y}{2}}
\newcommand{\Lxh}{\nicefrac{L_x}{2}}
\newcommand{\Lxih}{\nicefrac{L_{\xi}}{2}}

\newcommand{\ucds}{u_{\rm cds}}

\newcommand{\R}{\mathbb{R}}

\newcommand{\Lcr}{L_\textrm{cr}}
\newcommand{\Lth}{L_\textrm{th}}

\newcommand{\cmax}{c_\textrm{max}}
\newcommand{\eps}{\varepsilon}
\newcommand{\bx}{{\bf x}}
\newcommand{\ub}{u_{\mathrm{b},\pm}}
\newcommand{\ubr}{u_{\mathrm{b}}}
\newcommand{\sn}{\mathrm{sn}}
\newcommand{\psia}{\psi_{\mathrm{approx.}}}
\newcommand{\ud}{\mathrm{d}}
\newcommand{\ie}{\textit{i.e.}}
\newcommand{\eg}{\textit{e.g.}}

\definecolor{oneblue}{rgb}{0,0.0,0.75}
\usepackage[bookmarks,urlcolor=blue,citecolor=blue,linkcolor=blue,
                pagecolor=blue,colorlinks,hyperfigures]{hyperref}

 \hypersetup{
   hyperindex=true,
   pdftitle={}, pdfstartview=Fit,  pdfhighlight=/P,
   bookmarksnumbered=false, bookmarksopen=false,
   colorlinks=true, urlcolor=oneblue, linkcolor=oneblue, citecolor=oneblue,
   plainpages=false, pageanchor=true, pagebackref=true
 }

\begin{document}


\title{Onset of transverse instabilities of confined dark solitons}

 \author{M.~A.~Hoefer}
\affiliation{Department of Applied Mathematics, University of Colorado, Boulder, CO 80309, USA}
\email{hoefer@colorado.edu}
\author{B.~Ilan}
\affiliation{School of Natural Sciences, University of California - Merced, Merced, CA, 95343, USA}
\email{bilan@ucmerced.edu}

\date{\today}

\begin{abstract}
  We investigate propagating dark soliton solutions of the
  two-dimensional defocusing nonlinear Schr\"odinger /
  Gross-Pitaevskii (NLS/GP) equation that are transversely confined to
  propagate in an infinitely long channel.  Families of single,
  vortex, and multi-lobed solitons are computed using a
  spectrally-accurate numerical scheme.  The multi-lobed solitons are
  unstable to small transverse perturbations.  However, the
  single-lobed solitons are stable if they are sufficiently confined
  along the transverse direction, which explains their effective
  one-dimensional dynamics.  The emergence of a transverse
  modulational instability is characterized in terms of a spectral
  bifurcation.  The critical confinement width for this bifurcation is
  found to coincide with the existence of a propagating vortex
  solution and the onset of a ``snaking'' instability in the dark
  soliton dynamics that, in turn, give rise to vortex or multi-vortex
  excitations.  These results shed light on the superfluidic
  hydrodynamics of dispersive shock waves in Bose-Einstein condensates
  and nonlinear optics.
\end{abstract}


\maketitle


Solitary waves are ubiquitous in nonlinear dispersive systems.
Isolated solitary waves are often either elevation or depression
waves, also called bright or dark solitons, respectively.  Dark
solitons, also known as kink solitons, require a non vanishing
background and their mathematical study has been extensive
(cf.~\cite{Kivshar-00,Frantzeskakis-JPA-10,Kevrekidis_2015}).
Experimentally, temporal and spatial dark solitons have been observed
in optical fibers~\cite{Emplit-OC-87,Weiner-PRL-88,Krokel-PRL-88} and
waveguide arrays~\cite{Smirnov-PRE-06}, surface water
waves~\cite{Denardo-PRL-90,Chabchoub-PRL-13},
plasmas~\cite{Heidemann-PRL-09}, and Bose-Einstein condensates
(BECs)~\cite{burger_dark_1999,becker_oscillations_2008}.  Dark
solitons are one-dimensional objects that can naturally be extended to
higher dimensions as planar dark solitons.  
In experiments, the transverse direction is spatially confined.  For
example, by use of an appropriately shaped electromagnetic potential,
a BEC can be confined into a ``pancake'' or ``cigar'' shape, yielding
effective two-dimensional or one-dimensional dynamics, respectively.
Moreover, it has been observed experimentally that these
effectively one-dimensional BEC dynamics are stable when sufficiently
confined~\cite{burger_dark_1999,becker_oscillations_2008} and 
unstable otherwise~\cite{dutton_observation_2001,anderson_watching_2001}.  This
raises the question: what is the threshold confinement width for
effectively lower-dimensional dynamics of dark solitons in nonlinear
wave systems?

To address this question, we consider the mean field, superfluidic
(dissipationless) dynamics of a BEC governed by the (2+1)-dimensional
defocusing nonlinear Schr\"odinger / Gross-Pitaveskii (NLS/GP)
equation.  In one dimension, this equation admits dark soliton
solutions.  In multiple dimensions, the NLS/GP equation admits line
or planar dark solitons, which are uniform along all but the direction
of propagation.  Exact propagating line dark solitons solutions can be
written explicitly and their stability has been analyzed extensively.
In particular, they are modulationally unstable to small transverse
perturbations~\cite{kuznetsov_instability_1988}.

Previous studies have obtained critical confinement thresholds for
stationary or ``black'' solitons~\cite{Brand-PRA-02,Kevrekidis-04} and
propagating or ``gray'' solitons~\cite{muryshev_dynamics_2002}.  In
the latter case for gray solitons, the stability of approximate
soliton solutions of a (3+1)-dimensional GP equation with transverse
confinement achieved by a harmonic potential were obtained as follows.
An initial black soliton was dynamically evolved in the presence of
dissipation, leading to a reduction in amplitude and an approximate
confined gray soliton.  These solutions were then used to linearize
the GP equation and the resulting equations were evolved numerically
and analyzed for amplitude growth (instability).  

In our study, the critical confinement width for (2+1)-dimensional
propagating dark solitons is computed and found to occur at the onset
of a transverse instability.  Specifically, generalized bound state
(or solitary wave) solutions are computed in a two-dimensional
channel, which is confined along the $y$ axis and unbounded along the
$x$ axis.  Along the confined direction, either Dirichlet
(impenetrable wall) or Neumann (zero flux) boundary conditions are
specified.  In the $x$ direction, the computed bound states limit to
the one-dimensional transverse ground state with the possibility of a
phase jump from $-\infty$ to $\infty$.  We call the bound states with
Dirichlet boundary conditions confined dark solitons (CDSs).  Unlike
line dark solitons on the unbounded domain $\R^2$, or those satisfying
Neumann boundary conditions, for the case of impenetrable walls there
are no known CDS solutions in analytical form.  The CDSs are computed
using a spectrally accurate quasi-Newton fixed-point iterative scheme.
To accurately analyze the onset of the instability, the NLS/GP
equation is linearized around the CDS.  The eigenvalues of the
linearized equation are found to bifurcate from the origin at a
critical confinement width.  In particular, when the domain is
sufficiently narrow, all eigenvalues are purely imaginary (stable).
As the confinement width increases, two purely imaginary eigenvalues
coalesce at the origin and emerge as two real (unstable) eigenvalues
of opposite signs, whose corresponding eigenvectors are antisymmetric
along the transverse direction.  Dynamically, these eigenvectors
induce a transverse ``shear'' that breaks the bound state apart during
propagation.  This phenomenon, known as a ``snaking'' instability, has
been shown to give rise to single and multi-vortex
excitations~\cite{Brand-PRA-02,Kevrekidis_2015}.  The critical
confinement width obtained at the bifurcation of the discrete spectrum
is computed and characterized as a function of the soliton's
propagation speed and the type of boundary conditions.  We also
observe a new bound state solution branch bifurcating from the CDS
branch precisely at the critical confinement width.  These solutions
correspond to single propagating vortices.  Using direct numerical
simulations of the (2+1)-dimensional NLS/GP equation, we show that the
critical confinement width coincides with the onset of the transverse
instability regime of CDSs.  We also show that the $n$-lobed CDSs, $n
> 1$, exist only for sufficiently wide confinement.  These $n$-lobed
CDSs are always transversely unstable, leading to the generation of
vortices.

We remark that a background flow is present naturally in a shock wave.
In particular, dark solitons are intimately related to dispersive
shock waves (DSWs), also referred to as collisionless shock waves,
dissipationless shock waves, and undular bores~\cite{el_dispersive_2016}.  
A DSW connects two regions of a flow that
possess different parameters, such as density and velocity.  Unlike
viscous (dissipative) shock waves, a DSW consists of a modulated
wavetrain of oscillations.  In their seminal work, Gurevich and
Pitaevskii developed an asymptotic theory for DSWs, which they used to
show that the largest amplitude oscillation of a DSW can be well
approximated by a soliton~\cite{gurevich_nonstationary_1974}.  This
work was later extended to the NLS/GP equation where an approximate
dark soliton coincides with the slowest edge of the 
DSW~\cite{gurevich_dissipationless_1987}.  In addition to solitons, DSWs
have been observed experimentally in the aforementioned physical
systems as well (cf.~\cite{el_dispersive_2016}).  Moreover, the
multi-dimensional stability of DSWs has been connected to the
stability of the soliton edge~\cite{Hoefer-MMS-12}. Thus, the results
of this study shed light on the dynamics of dark solitons and DSWs.

\section{Problem formulation}
\label{sec:problem-formulation}

We consider the complex field $\psi(\bx,t)$ whose dynamics evolve
according to the defocusing (repulsive) NLS/GP equation.  In
dimensionless form this is
\begin{equation}
  \label{eq:1}
    i \psi_t + \frac{1}{2} \Delta \psi  - |\psi|^2 \psi = 0~, 
    ~~ \Delta \doteq \partial^2_{xx} + \partial^2_{yy}~.
\end{equation}
This equation governs the mean field dynamics of a BEC wavefunction,
the electric field dynamics of intense laser beams in optical Kerr
media, and other nonlinear wave systems.  Here, $t>0$ is time,
$\bx=(x,y)$, where $x\in\R$ is the background flow direction, and
$y\in(-\Lyh,\Lyh)$, which corresponds to an infinitely long 2D channel
of width $L_y$.  In what follows, we are interested in two kinds of
boundary conditions along the transverse direction.
\begin{enumerate}
\item
  Zero-flux walls  (Neumann boundary conditions)
\begin{subequations}
  \label{eq:BC-trans}
  \begin{equation}
    \label{eq:Neumann}
   \psi_y(x,\pm\Lyh,t) = 0~.
 \end{equation}
\item 
 Impenetrable walls (Dirichlet boundary conditions)
  \begin{equation}
    \label{eq:Dirichlet}
    \psi(x,\pm\Lyh,t) = 0~.
  \end{equation}
\end{subequations}
\end{enumerate}

\subsection{Invariant quantities}

It is expedient to recap some of the well known invariances of~\eqref{eq:1}.
\begin{itemize}
\item
  Galilean invariance with velocity $v\in R$,
  \begin{equation}
    \label{eq:Gal}
    \psi'(x',y,t) = \psi(x' - vt,y,t) e^{i(vx' - \nicefrac{v^2 t}{2})}.
  \end{equation}
\item
Dilation invariance with amplitude $a\in R$,
\begin{equation}
  \label{eq:dilation}
  \psi'(x',y',t') = a \psi(a x', a y',a^2 t').
\end{equation}
\item 
Phase invariance with angle $\theta\in [0,2\pi)$,
\begin{equation}
  \label{eq:phase}
  \psi'(x',y',t') = e^{i\theta} \psi(x',y',t').
\end{equation}
\end{itemize}
These invariances are useful for prescribing simplified boundary
conditions of the bound state solutions.

As described below, there are  conserved quantities that are related to these 
invariant quantities. However, due to the non-vanishing boundary conditions, 
the conserved quantities are non-standard. To find them, we first need to 
define the bound state problem.

\subsection{Bound states and background states}

Given a confinement width $L_y$, we seek a one-parameter family of
confined  dark soliton solutions of~\eqref{eq:1} in the form
\begin{equation}
  \label{eq:4}
  \begin{split}
    \psi(x,y,t;\mu) = \ucds(\xi,y;\mu) e^{-i \mu t}, \quad \xi = x - ct,
  \end{split}
\end{equation}
where  the CDS profile, $\ucds(\cdot)$, is a complex-valued bound state,
$c \ge 0$ is the soliton speed, and $\mu$
is the frequency (also called propagation constant or 
chemical potential~\cite{pethick_bose-einstein_2008}).
When $c=0$, the confined dark soliton is stationary (``black''). 
When $c>0$, the confined dark soliton is propagating (``gray'').
Inserting the ansatz~\eqref{eq:4} into~\eqref{eq:1} gives the 
nonlinear boundary value problem 
\begin{equation}
  \label{eq:u}
  \begin{split}
    \mu \ucds + \frac{1}{2} \Delta \ucds - i c \partial_x u_{{\rm
        cds}}  - |\ucds|^2 \ucds = 0,  
  \end{split}
\end{equation}
subject to either of the transverse boundary conditions~\eqref{eq:BC-trans}.
The boundary conditions for the CDS along the flow direction, \ie, the
$x$ axis, are prescribed below.

Using~\eqref{eq:Gal}, we may assume, by Galilean invariance
\eqref{eq:Gal}, that as $x\to\infty$, the background (far field) flow
is stationary.  Hence, the speed $c$ plays no role in determining the
background flow.  Therefore, we require that the bound state
approaches a {\bf background state} as $x\to \pm \infty$, \ie, the
{\bf background boundary condition}
\begin{equation}
  \label{eq:ub1}
  \ucds(x,y) \stackrel{x\to\pm \infty}{\longrightarrow} \ub(y)~, 
\end{equation}
where $\ub(y)$ satisfies the associated ordinary differential equation
\begin{equation}
  \label{eq:ub}
  \frac{1}{2}\ \ub'' + \mu \ub - |\ub|^2\ub = 0~
\end{equation}
and the same  transverse boundary conditions~\eqref{eq:BC-trans} as $u(x,y)$.
Here, the {\bf background state} $\ub(y)$ is unique up to a complex phase.
In fact, the complex phase of $u(x,y)$ varies along the $x$ direction
and can approach different values as $x\to -\infty$ and  $x\to +\infty$.
For convenience, we also define 
$$
\ubr(y) \doteq |\ub(y)|~.
$$
In addition, using~\eqref{eq:dilation}, we can normalize the density
of the background state along the axis of symmetry as
\begin{equation}
  \label{eq:normalize}
  \ubr(0) = 1~.
\end{equation}
Physically, the background state corresponds to a dispersive
(inviscid) generalization of Poiseuille flow for a compressible fluid.
We note that the corresponding boundary value problem for the
one-dimensional NLS/GP equation was studied in~\cite{Carr-PRA-00}.

\subsection{Particle number}

We consider the initial value problem~\eqref{eq:1}, subject to either
of the transverse boundary conditions~\eqref{eq:BC-trans} and the
background boundary condition
\begin{equation}
  \label{eq:psib}
  \psi(x,y,t) \stackrel{ x \to \pm \infty}{\longrightarrow} \ub(y)
  e^{-i\mu t}~,
\end{equation}
where $\ub$ is the background state defined above.  We call a solution
of this problem a confined dark soliton (CDS).  As we shall see, CDS
solutions decay rapidly to the background state.  Therefore, we
define the {\bf particle number} (in analogy with its interpretation
in BECs) as
\begin{equation}
  \label{eq:N}
  \mathcal{N}[\psi] = \int_{-\infty}^\infty \int_{-\Lyh}^{\Lyh}
  \left[\ubr^2(y) - | \psi(x,y,t) |^2 \right] \,  \ud y\, \ud x~.
\end{equation}
This quantity is finite and positive for CDSs.  That this quantity is
conserved in time follows from taking its time derivative,
using~\eqref{eq:1}, integration by parts, and the boundary conditions.

\section{Line dark soliton with zero-flux walls}
\label{sec:Neumann}

For the case of zero-flux walls~\eqref{eq:Neumann}, the background is
simply constant regardless of the confinement width.  It follows
from~\eqref{eq:normalize}, \eqref{eq:ub}, and~\eqref{eq:u} that
$\mu=1$ and $\ubr(y) \equiv 1$.  Therefore, the {\bf exact} solution
takes the form of a line dark soliton uniform in $y$, \ie,
\begin{equation}
  \label{eq:10}
  \begin{split}
    \psi(x,y,t;c) &= \left \{ i c + \nu \tanh [ \nu(x-ct)] \right \} e^{-i  t}, \\
    \nu^2 + c^2 &= 1, \quad 0 \le c < 1 ,
  \end{split}
\end{equation}
up to an overall phase.  It is well known that this dark soliton
solution exhibits an instability for sufficiently long wavenumber
perturbations in the transverse
direction~\cite{kuznetsov_instability_1988}.  It follows from the
critical transverse wavenumber (see~\cite{kuznetsov_instability_1988})
that, in this case, the solution with zero-flux walls is transversely
unstable when
\begin{equation}
  \label{eq:15}
  L_y > \Lcr(c) = \frac{\pi}{\sqrt{-1-c^2+2\sqrt{1-c^2+c^4}}}~.
\end{equation}
It follows that for dark solitons with zero-flux walls, as the speed
increases from 0 to 1, $\Lcr(c)$ increases monotonically from $\pi$ to
$\infty$.  Loosely speaking, this means that, as the speed increases,
the smaller amplitude line dark soliton becomes more stable to
transverse perturbations.  This feature has been observed for
transverse harmonic potential confinement in (3+1)-dimensions
\cite{muryshev_dynamics_2002} and we will see similar behavior for the
impenetrable wall boundary condition in what follows.

\section{CDS  with impenetrable walls}

In many physical systems, transverse boundary conditions can be well
approximated by impenetrable walls (Dirichlet boundary
conditions)~\eqref{eq:Dirichlet}.  In this case, there is no exact
analytic CDS solution and there is no known exact conditions for
stability analogous to~\eqref{eq:15}.  However, these CDS solutions
can be computed.  To do so, it is helpful to obtain approximate
analytical solutions as follows.  This is achieved below in two steps:
first we find the background states and then use them to construct
approximate CDSs by an approximate ``nonlinear separation of
variables'', a technique well-known in the BEC community
\cite{Kevrekidis_2015}.

\begin{figure}
  \centering 
  \includegraphics{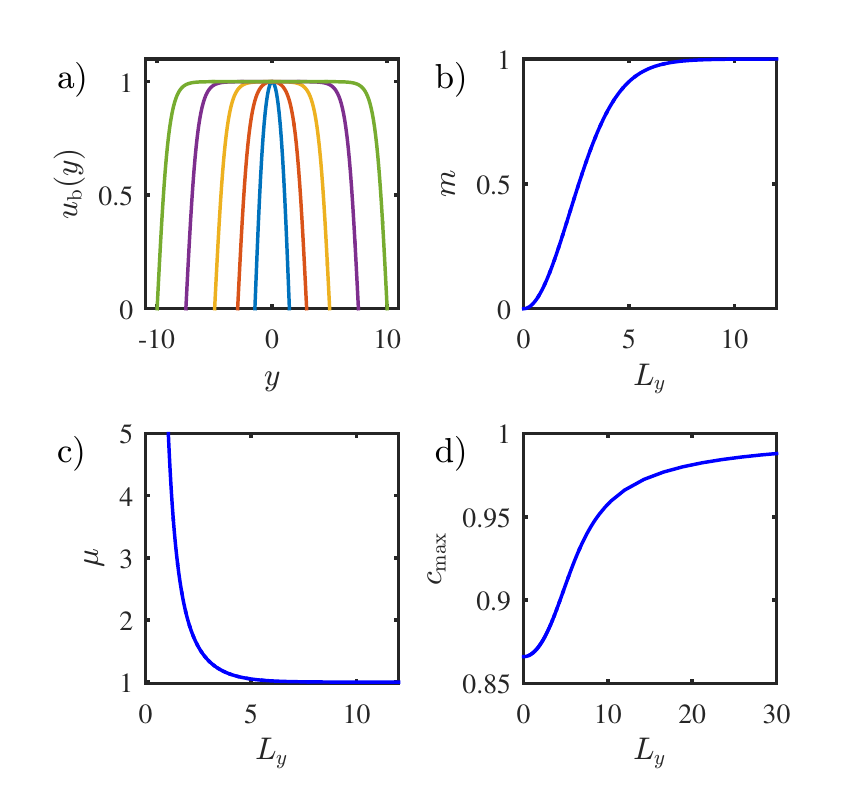}
  \label{fig:background_properties}
  \caption{(Color online) Normalized background states
    [Eq.~\eqref{eq:sn}] and their salient features.  a) Multiple
    transverse profiles for different widths $L_y$.  b) Elliptic
    parameter $m$ [Eq.~\eqref{eq:m}].  c) Frequency $\mu$
    [Eq.~\eqref{eq:sn}].  d) Maximum approximate CDS speed
    [Eq.~\eqref{eq:2}].}
\end{figure}

There is a one-parameter family of solutions of the background
state~\eqref{eq:ub} with boundary conditions~\eqref{eq:Dirichlet} that
can be expressed in terms of a Jacobi elliptic function as
\begin{equation}
  \label{eq:sn}
  \begin{split}
    \ubr(y;L_y) &= \sn\, [\kappa(y +\Lyh),m ], \\*[2mm]
    \kappa &= \frac{1}{\sqrt{m}}, \quad \mu = \frac{1+m}{2 m} .
  \end{split}
\end{equation}
The elliptic parameter $0 < m(L_y) < 1$ is determined from the boundary
conditions according to
\begin{equation}
  \label{eq:m}
  L_y = 2 \sqrt{m} K(m),
\end{equation}
where $K(m)$ is the complete elliptic integral of the first kind.
Thus, given $L_y$, we compute $m$ using~\eqref{eq:m} and find
$\mu,\kappa$, and $\ubr$ using~\eqref{eq:sn}.
Figure~\ref{fig:background_properties} depicts several such background
states as well as the elliptic parameter, frequency, and maximal
velocity (see below) for varying confinement widths.  In particular,
Fig.~\ref{fig:background_properties}(c) shows that as the confinement
width increases, the frequency converges rapidly to its unconfined
value, which is the same as that of Neumann (constant
flux) transverse boundary conditions, \ie, $\lim_{L_y\to\infty} \mu
=1$. Note that in this limit, formally, $m\to1$ and $\sn(\cdot;m)\to
\tanh(\cdot)$. However, this is a singular limit and, in fact, the
limiting $\ubr$ is constant along the $y$ direction.

In addition, it can be shown from~\eqref{eq:sn} that
\begin{equation}
  \label{eq:9}
  \int_{-\Lyh}^{\Lyh} \ubr^2(y)\, \ud y = \frac{2}{\sqrt{m}}
  \left[K(m) - E(m)\right], 
\end{equation}
where $E(m)$ is the complete elliptic integral of the second kind.
These relations are useful for the subsequent analysis and
computations.

\subsection{Approximate analytical CDS with impenetrable walls}
\label{sec:approximate-cds}

Because the NLS/GP equation is nonlinear, strictly speaking, linear
methods for partial differential equations, such as the method of
separation of variables, do not apply for this equation.  In spite of
this, it is shown below that one can obtain an approximate separable
solution, which, in turn, is found to be fairly accurate and useful.
Such an approach has been shown to work well for a variety of
transverse configurations, including tight harmonic confinement
\cite{Kevrekidis_2015}.  Intuitively, for sufficiently tight
confinement, the energy to excite transverse dynamics is large and one
expects the transverse direction to be ``frozen''.

Thus, we seek a solution of~\eqref{eq:1}  and~\eqref{eq:Dirichlet} 
in the separable form
\begin{equation}
  \label{eq:14}
  \psi(x,y,t;c,L_y) = \ubr(y;L_y) \tilde{\psi}(x,t;c)~,
\end{equation}
where $\ubr$ is the background solution~\eqref{eq:sn}.
Inserting this ansatz into~\eqref{eq:1}, 
multiplying by $\ubr$, 
and integrating over $y$, yields
\begin{equation}
  \label{eq:16}
  i \frac{\partial \tilde{\psi}}{\partial t} 
  + \frac{1}{2}\frac{\partial^2 \tilde{\psi}}{\partial x^2}  
  - \alpha |\tilde{\psi} |^2 \tilde{\psi}  = 0,
\end{equation}
where the nonlinear coefficient is determined through the confinement
width $L_y$ as 
\begin{equation}
  \label{eq:26}
  \begin{split}
    \alpha(L_y) &= \frac{\int_0^{K(m)} \sn^4(y,m) \, dy}{\int_0^{K(m)} \sn^2(y,m) \, dy} \\
    &= \frac{1}{3} \left [ 1 + \frac{2}{m} + \frac{E(m)}{E(m)-K(m)}\right ] .
  \end{split}
\end{equation}
It follows from~\eqref{eq:26} that
\begin{equation}
  \label{eq:3}
  \lim_{L_y \to 0} \alpha = \lim_{m \to 0} \alpha = \frac{3}{4} .
\end{equation}
The physical meaning of $\alpha$ is an effective nonlinear coupling
constant (or scattering length in BECs).  In the unconfined case,
$\nu^2+c^2 = \alpha^2 =1$. In the confined case, $\alpha <1$, \ie,
there is a reduced speed that a dark soliton can have due to the
confinement, which has been observed experimentally in BECs
\cite{becker_oscillations_2008}.

Equation \eqref{eq:16} admits dark soliton solutions that we use in
order to construct a family of {\bf approximate CDS solutions}
propagating with speed $c$ as
\begin{equation}
  \label{eq:19}
  \begin{split}
    \psia&(x,y,t) = \ubr(y) \frac{ i c + \nu \tanh\left [\nu\xi
      \right ]}{\sqrt{\alpha}} e^{-i
      \mu t}, \\
    \nu &= \sqrt{\alpha - c^2}, \quad \mu =
    \alpha, \quad 0 \le c < \sqrt{\alpha} . 
  \end{split}
\end{equation}
We note that the approximate CDS~\eqref{eq:19} does not satisfy the
NLS/GP equation exactly, though it does satisfy the boundary
conditions \eqref{eq:Dirichlet}, \eqref{eq:ub1}, and
\eqref{eq:normalize}.  The far-field behavior
\begin{equation}
  \label{eq:5}
  \lim_{x \to \pm \infty} \psia(x,0,t) = \frac{i c \pm
    \nu}{\sqrt{\alpha}} e^{-i \mu t},
\end{equation}
implies the phase jump
\begin{equation}
  \label{eq:13}
  \Delta \phi = \pi - 2\tan^{-1}(c/\sqrt{\alpha - c^2})
\end{equation}
as the dark soliton is traversed.  Furthermore, it follows
from~\eqref{eq:19} that for a given confinement width, $L_y$, there is
a maximal speed given by
\begin{equation}
  \label{eq:2}
  \cmax (L_y)= \sqrt{\alpha (L_y)}~.
\end{equation}
Figure~\ref{fig:background_properties}(c) shows the variation of this maximal speed 
with the confinement width. 
As $c \to \cmax (L_y)$, the CDS~\eqref{eq:19}
approaches the uniform in-$x$ background state, $u_\mathrm{b}(y)$.
It follows from~\eqref{eq:3} that 
$$
\nicefrac{\sqrt{3}}{2} \le \cmax < 1~.
$$
We also find the particle number~\eqref{eq:N} for the approximate CDS as
\begin{equation}
  \label{eq:33}
  \mathcal{N}[\psia] = \frac{4}{\alpha \sqrt{m}} \left [ K(m) - E(m) \right ] \sqrt{\alpha - c^2} .
\end{equation}

We remark that there are other approaches in the literature to obtain
approximate solutions, \eg, the Lagrangian approach
\cite{Kevrekidis_2015}. In all cases, the approximation cannot satisfy
exactly both the NLS/GP equation and the boundary conditions.  The
method above was chosen because it satisfies the boundary conditions
exactly.  The computations discussed below verify the utility of this
approach.

\subsection{CDS computations}
\label{sec:CDS_computations}

\begin{figure}
  \centering 
  \includegraphics[scale=0.25]{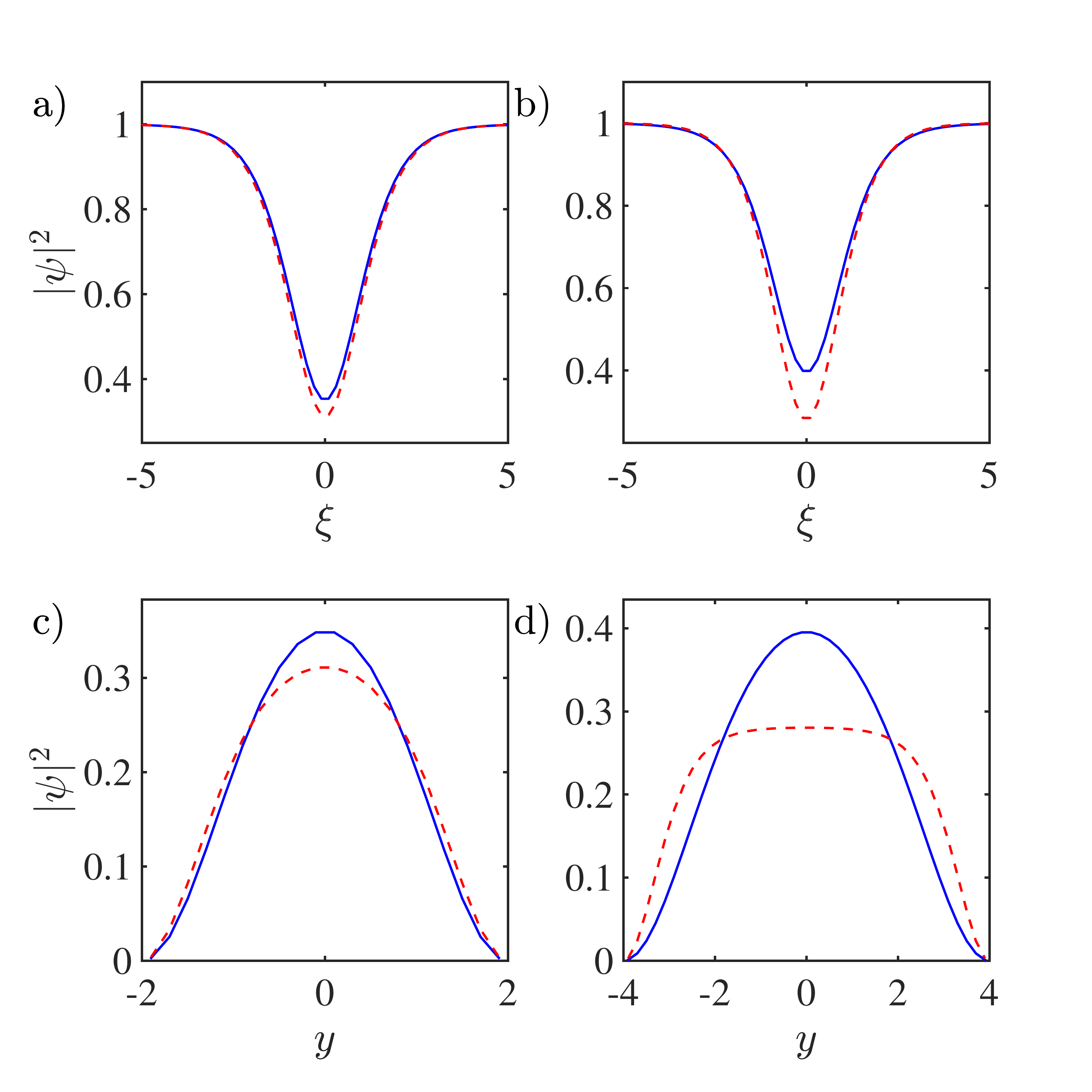}
  \caption{(Color online) Comparison of approximate (dashed) and exact
    (solid) CDS cross sections for $c = 0.5$.  (a,b) Longitudinal
    cross sections along $y = 0$ with transverse confinement widths $L_y
    = 4$ and $L_y = 8$, respectively.  (c,d) Transverse
    cross sections along $\xi = 0$ with transverse confinement widths $L_y
    = 4$ and $L_y = 8$, respectively.}
  \label{fig:gray_c_5_L_3_compare}
\end{figure}

\begin{figure}
  \centering 
  \includegraphics{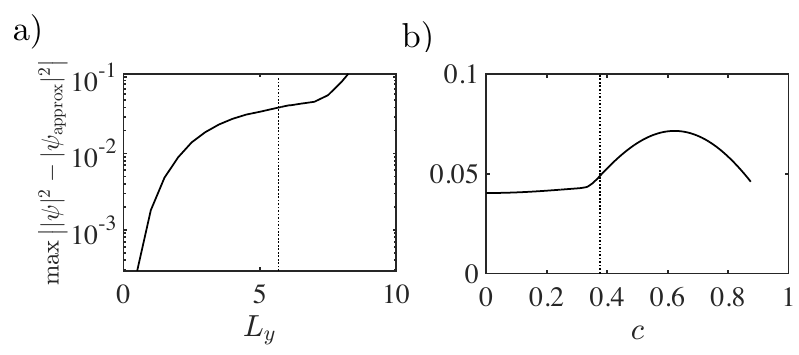}
  \caption{Maximum absolute error in density $|\psi|^2$ between
    approximate and numerical CDSs.  a) Fixed $c = 0.25$, variable
    width $L_y$. b) Fixed $L_y = 6$, variable speed $c$.  The dotted
    vertical line separates unstable and stable CDS solutions (see
    Sec.~\ref{sec:line-stab-analys}).}
  \label{fig:approx_error}
\end{figure}

\begin{figure}
  \centering 
  \includegraphics{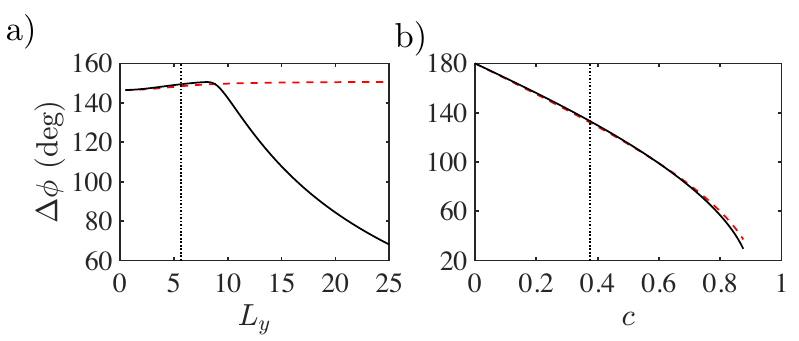}
  \caption{(Color online) Dependence of the phase jump across the CDS
    solution as a) the confinement width $L_y$ is varied with fixed
    CDS speed $c = 0.25$ and b) as the speed $c$ is varied with fixed
    width $L_y = 6$.  These plots show deviation from the approximate
    CDS phase jump (dashed) for large $L_y$.  The dotted vertical line
    separates unstable and stable CDS solutions (see
    Sec.~\ref{sec:line-stab-analys}).}
  \label{fig:cds_phase_c_L}
\end{figure}

The CDS solutions with impenetrable walls are computed using the
spectrally accurate quasi-Newton iterative method described in
Appendix~\ref{sec:numer-comp}.  The approximate analytic CDS described
above is used as an initial guess for these iterations.  Simple
continuation of the computed CDS solutions is used for larger $L_y$.
Figure~\ref{fig:gray_c_5_L_3_compare} presents a comparison of
cross-sections between an approximate CDS and the numerically computed
one, demonstrating good agreement for sufficiently tight confinement.
This is further confirmed in Fig.~\ref{fig:approx_error}a where the
error between the approximate and numerical CDS densities for a fixed
speed is shown to decrease with decreasing $L_y$.
Figure~\ref{fig:approx_error}b gives the error for fixed width and
variable speed.
In addition, Fig.~\ref{fig:cds_phase_c_L} shows that the phase jump
across the approximate CDS is fairly accurate compared to the
numerically computed CDS for a wide range of confinement widths and
propagation speeds.  This underscores the utility of the approximate
CDS.

\begin{figure}
  \centering 
  \includegraphics[scale=0.25]{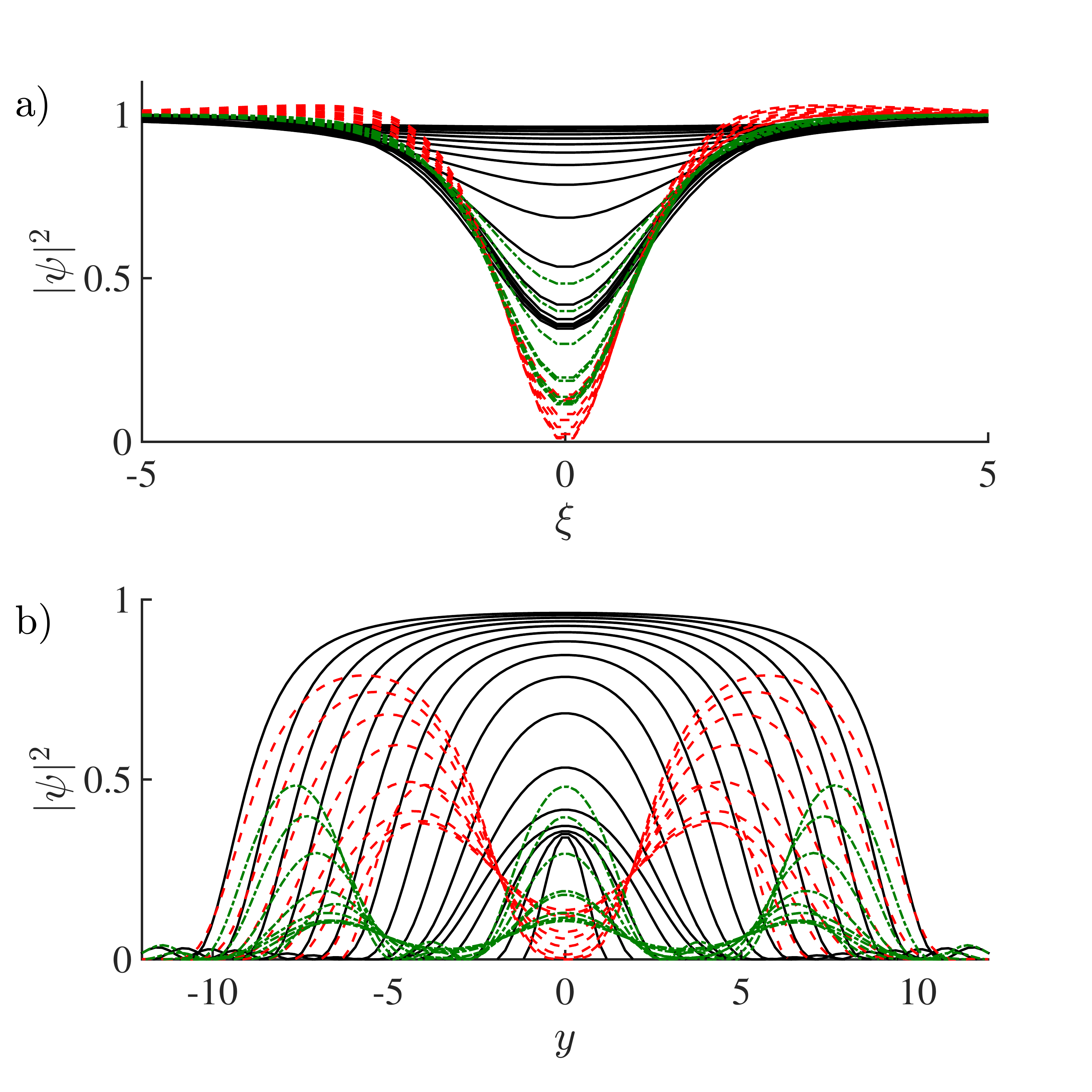}
  \caption{(Color online) Cross sections along a) the longitudinal
    direction for $y=0$ and b) along the transverse direction with
    $\xi=0$ of single-lobed (solid black), double-lobed (dashed red),
    and triple-lobed (dash-dotted green) CDSs with propagation speed
    $c = 0.5$ and a range of confinement widths $L_y$.}
  \label{fig:multi_lobed_ds}
\end{figure}

In addition, CDS solutions can possess any number of oscillation lobes
within the central depression.  Figure~\ref{fig:multi_lobed_ds}
presents examples of gray (propagating) solitons with speed $c=0.5$
for varying confinement widths, which possess one, two, and three
lobes.  The maximal possible number of lobes depends on $c$ and $L_y$.
The entire family of 1-lobe CDSs can be computed using a continuation
method in $L_y$ -- these CDSs bifurcate from the uniform background
state $\ubr(y)$.  For $n>1$, we find that an $n$-lobed CDS exists if
the confinement width $L_y$ is above a critical value, $\Lth^n(c)$.
When the Newton solver is initialized with the approximate (1-lobe)
CDS \eqref{eq:19} for $\Lth^n(c) < L_y < \Lth^{n+1}(c)$, the iteration
converges to an $n$-lobe CDS.  Continuation along each $n$-lobe branch
enables the numerical determination of $\Lth^n(c)$.  For example,
$\Lth^2(0.5) \approx 12.8$ and $\Lth^3(0.5) \approx 21.8$.  We
interpret these lobed solutions as ``nonlinear excited states'' of the
channel.






\subsection{Bifurcation of CDS families and vortex solitons}

\begin{figure*}
  \centering 
  \includegraphics{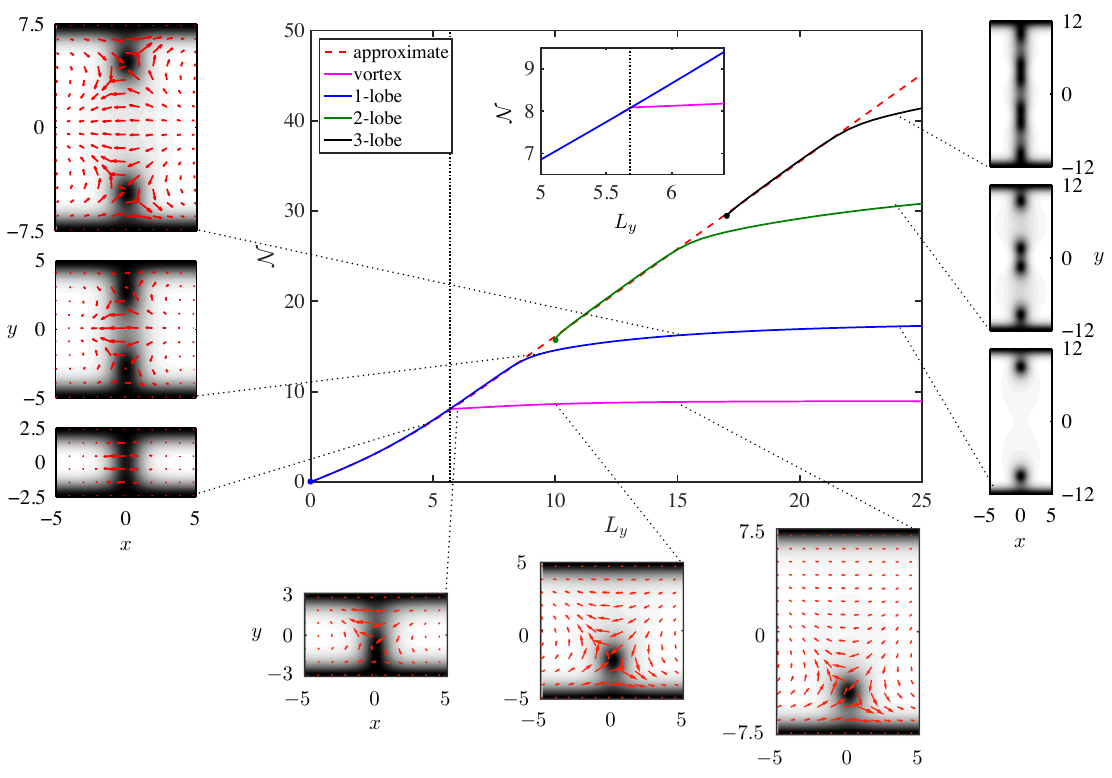}
  \caption{(Color online) Phase diagram depicting the conserved
    quantity $\mathcal{N}$ and contour plots of 1, 2, and 3-lobed
    channel CDS and solitonic vortex solutions with fixed $c = 0.25$
    as $L_y$ is varied.  The dotted vertical line corresponds to the
    critical confinement width $\Lcr \approx 5.68$ below which CDS
    solutions are stable (see Sec.~\ref{sec:line-stab-analys}).  The
    critical width corresponds to the minimal $L_y$ such that the
    solitonic vortex exists.}
  \label{fig:bifurcation_L}
\end{figure*}

To better understand the properties of the different CDS families,
Fig.~\ref{fig:bifurcation_L} depicts the particle number $\mathcal{N}$
for the 1-, 2-, and 3-lobed CDSs as well as for the vortex.  The
1-lobed CDSs bifurcate from the uniform background state for which
$\mathcal{N}[u_\mathrm{b}] = 0$, whereas, the $n$-lobed CDS bifurcates
from a suitably close approximate CDS (dashed red curve) with a
sufficiently large particle number.

In addition, our CDS computations reveal solitonic vortices.  Vortices
have been known and studied extensively for many years.  In most
cases, vortices arise in pairs of opposite circulation such that their
total angular momentum vanishes.  However, single solitonic vortices
have been predicted and studied theoretically
(cf.~\cite{Brand-PRA-02,Komineas-PRA-03,Ma-PRA-10,Middelkamp-PD-11,
  Aioi-PRL-11,Simula-PRA-11,Becker-NJP-13,Munoz-PRL-14,Scherpelz-14}).
Our study here focuses on two-dimensional vortices that have been
studied extensively in BECs (see, \eg, \cite{Kevrekidis_2015} and
references therein).  We note that the three-dimensional counterpart,
a vortex line, has been fairly elusive yet has been observed recently
in Fermi gases~\cite{Ku-PRL-14,Bulgac-PRL-14} and
BECs~\cite{Donadello-PRL-14}.  In addition to the aforementioned lobed
CDSs, Fig.~\ref{fig:bifurcation_L} presents the particle number for
solitonic vortex solutions with varying confinement widths.  We note
that the particle number of solitonic vortices is equal or lower than
for the 1-lobed CDS. This suggests that solitonic vortices should be
persistent structures, even for wide channels.

\begin{figure} [h!]
  \centering 
  \includegraphics{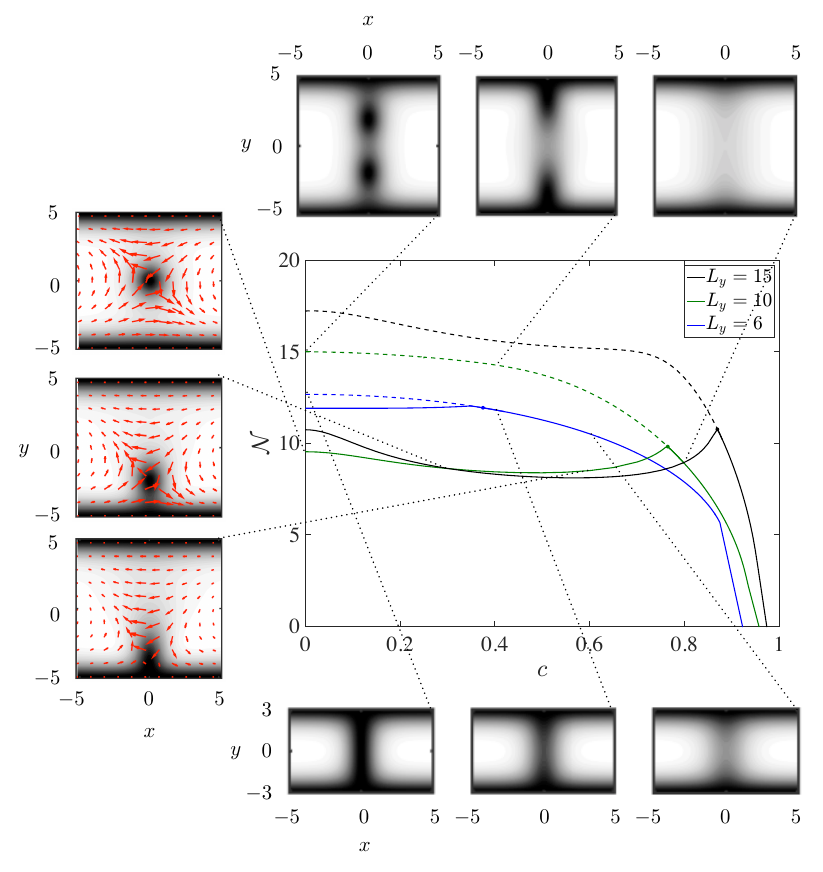}
  \caption{(Color online) Phase diagram depicting the conserved
    quantity $\mathcal{N}$ and contour plots of 1-lobed CDS and vortex
    solutions with variable speeds $c$ and several confinement widths
    $L_y$.  Filled circles correspond to the crossover from unstable
    (dashed curves) to stable (solid curves) CDS solutions (see
    Sec.~\ref{sec:line-stab-analys}).}
  \label{fig:bifurcation_c}
\end{figure}

In addition, Fig.~\ref{fig:bifurcation_c} presents the particle
numbers for 1-lobe CDSs and vortex solitons with varying propagation
speeds. In this figure, the stable and unstable modes are
distinguished by different curve types.  We now describe the stability
properties of these solutions.

\section{Stability analysis and direct numerical simulations}
\label{sec:line-stab-analys}

As mentioned in Section~\ref{sec:Neumann}, the critical confinement
width for transverse stability of line dark solitons with zero-flux walls
(Neumann boundary conditions) is given analytically by eq.~\eqref{eq:15}.
However, there is no such simple expression for CDS with impenetrable walls
(Dirichlet boundary conditions).  Here we investigate the linear (spectral)
transverse stability of the CDSs.  

\begin{figure}
  \centering 
  \includegraphics[width=\columnwidth]{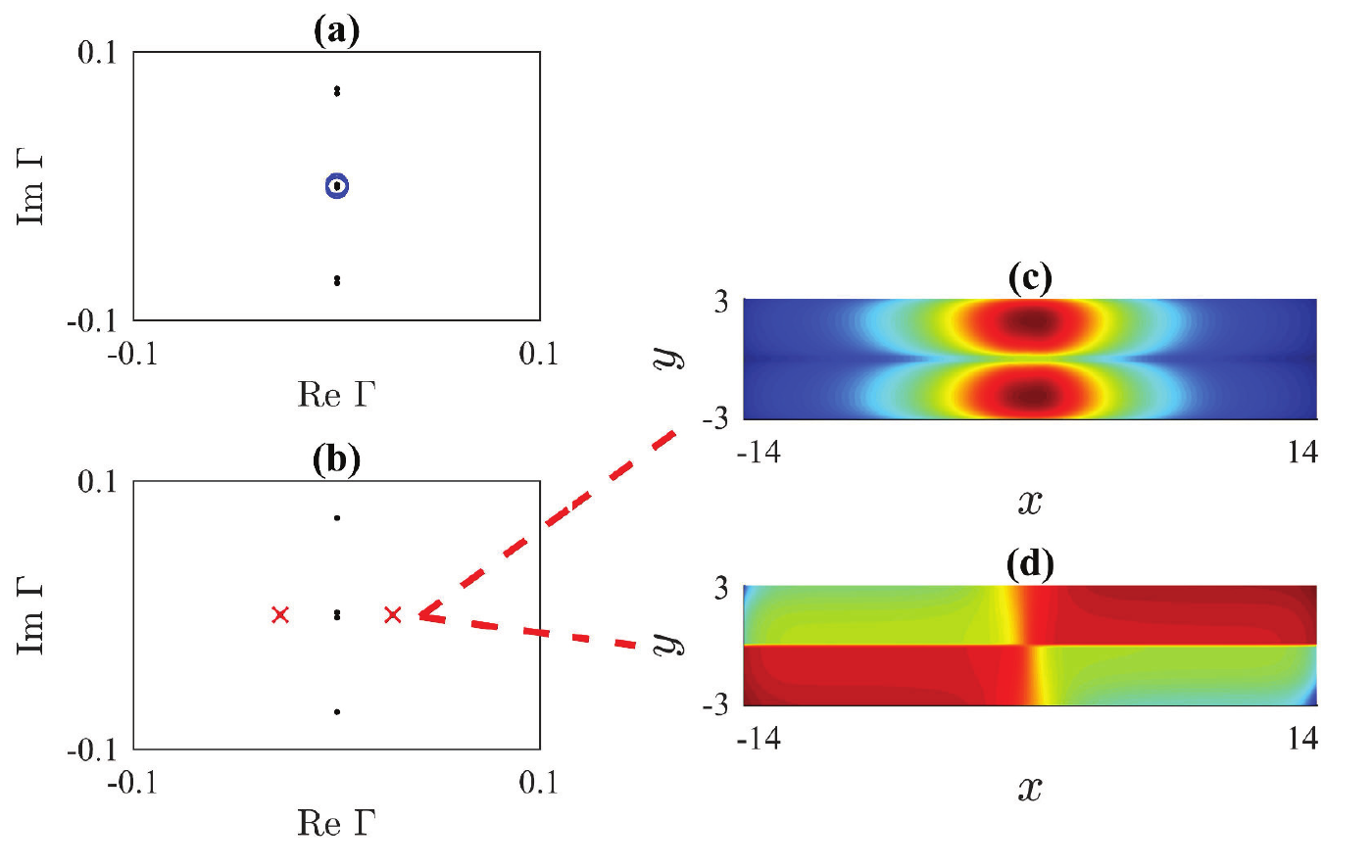}
  \caption{(Color online) A portion around the complex origin of the linearized
  spectra of a CDS with speed $c=0.5$ and confinement widths (a) $L_y
  = 6.4$ and (b) $L_y = 6.6$.  Two stable (purely imaginary)
  eigenvalues [denoted by $\varodot$ in (a) and almost
  indistinguishable] coalesce at the origin and emerge as unstable
  (real) eigenvalues [denoted by $\times$ in (b)].  Contour plots of
  the intensity (c) and phase (d) of one of the unstable
  eigenfunctions. }
  \label{fig:spectral_bifurcation_unstable_mode}
\end{figure}

In Appendix~\ref{sec:linearization} the linearized NLS/GP equation
around a CDS solution is derived (also termed the Bogoliubov-de Gennes
equations in BEC studies).  We compute the spectrum of the linearized
equation for varying confinement width $L_y$.  For a given soliton
speed, $c$, when the domain is sufficiently narrow, the spectrum of
the linearized operator consists of purely imaginary (stable)
discretized eigenvalues and continuous spectrum (radiation modes).
When the confinement width reaches a critical value, $L_y = \Lcr(c)$,
two purely imaginary eigenvalues coalesce at the origin.  For $L_y >
\Lcr(c)$, these eigenvalues emerge as two real (unstable) eigenvalues
of opposite signs, whose corresponding eigenvectors are antisymmetric
along the transverse direction.
Figure~\ref{fig:spectral_bifurcation_unstable_mode}[(a) and (b)]
depicts this bifurcation when $c=0.5$ for which $\Lcr\approx 6.55$.
Figure~\ref{fig:spectral_bifurcation_unstable_mode}(c) shows the
intensity and complex phase of one of the unstable
eigenfunctions. Whereas the intensity is symmetric, the complex phase
is odd in both $x$ and $y$, which gives rise to the ``snaking''
instability observed in the dynamics.

Our computations of the linearized spectrum reveal that the
1-lobed CDSs are stable when they are sufficiently confined, \ie, for
$L_y < \Lcr(c)$, where $\Lcr(c)$ is the critical confinement width.
Note that the background solution $u_b(y)$ is linearly stable.
Figure~\ref{fig:lcr_D_and_N} presents a plot of the critical
confinement width $\Lcr(c)$ for both Neumann and Dirichlet boundary
conditions.  In general, we find that
\begin{equation}
  \label{eq:17}
  \Lcr(c) \ge \Lcr(0) \approx 5.5 , \quad 0 \le c < 1 .
\end{equation}
The result by~\cite{Brand-PRA-02} for black solitons was that $\Lcr(0)
\approx 6$, which is close to the value we obtain using spectral
analysis.

The inset of Fig.~\ref{fig:bifurcation_L} shows that the solitonic
vortex solution branch bifurcates from the 1-lobe CDS solution branch
precisely at $\Lcr$.  We can therefore interpret the onset of the
1-lobe CDS instability as precisely the confinement width at which a
lower particle number solitonic vortex solution appears.  This
interpretation can also be corroborated by dynamical evolution of the
1-lobe CDS described below.

In Fig.~\ref{fig:bifurcation_c} above, the stable and unstable modes
were distinguished by different curve types.  For a given confinement
width, the modes become unstable when their particles number increases
beyond a critical threshold.  In general, the multi-lobed CDSs are
all unstable.
 
\begin{figure}
  \centering 
  \includegraphics[scale=0.8]{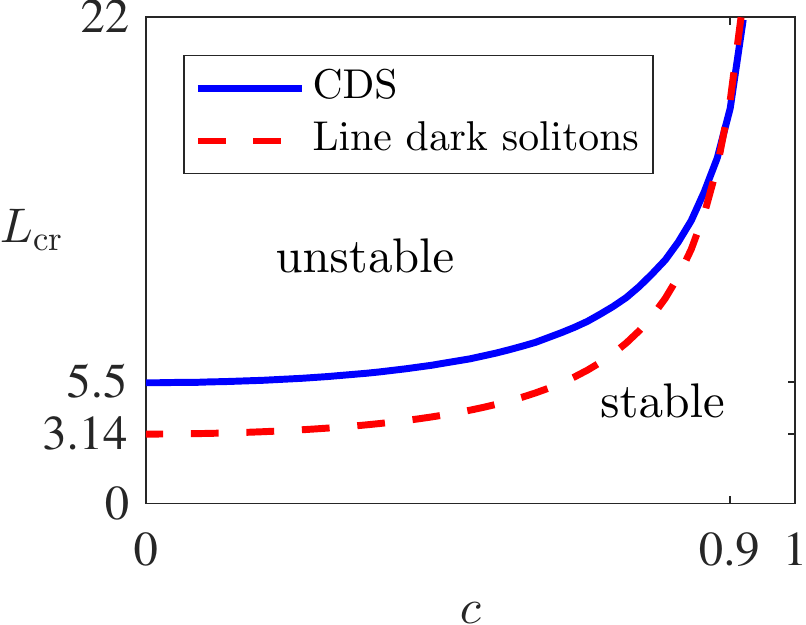}
  \caption{(Color online) 
    Critical confinement widths as functions of propagation speed, 
    $\Lcr(c)$, for line dark solitons with constant-flux walls (dashed) 
    and CDSs with impenetrable walls (solid).  
    For $L_y > \Lcr(c)$ the solutions are unstable.}
  \label{fig:lcr_CDS_and_line_solitons}
\end{figure}

Finally, to test the linear stability theory, we carry out direct
numerical simulations of the NLS/GP equation~\ref{eq:1} with
impenetrable walls and initial conditions that correspond to a CDS
with a small amount of ``noise''.  The numerical method is explained
in Appendix~\ref{sec:dynamical-evolution}.
Figure~\ref{fig:dynamics_c_25} presents some of the results, which
show that $\Lcr(c)$, obtained from spectral analysis, is indeed the
critical confinement width for nonlinear stability.  Moreover, these
simulations reveal that, when $L_y > \Lcr$, the CDS undergoes a
snaking instability and can break up into a single solitonic vortex as
in Fig.~(b) or counterpropagating solitonic vortex pairs as in
Fig.~(c).
\begin{figure*}
  \centering 
  \includegraphics[scale=0.2]{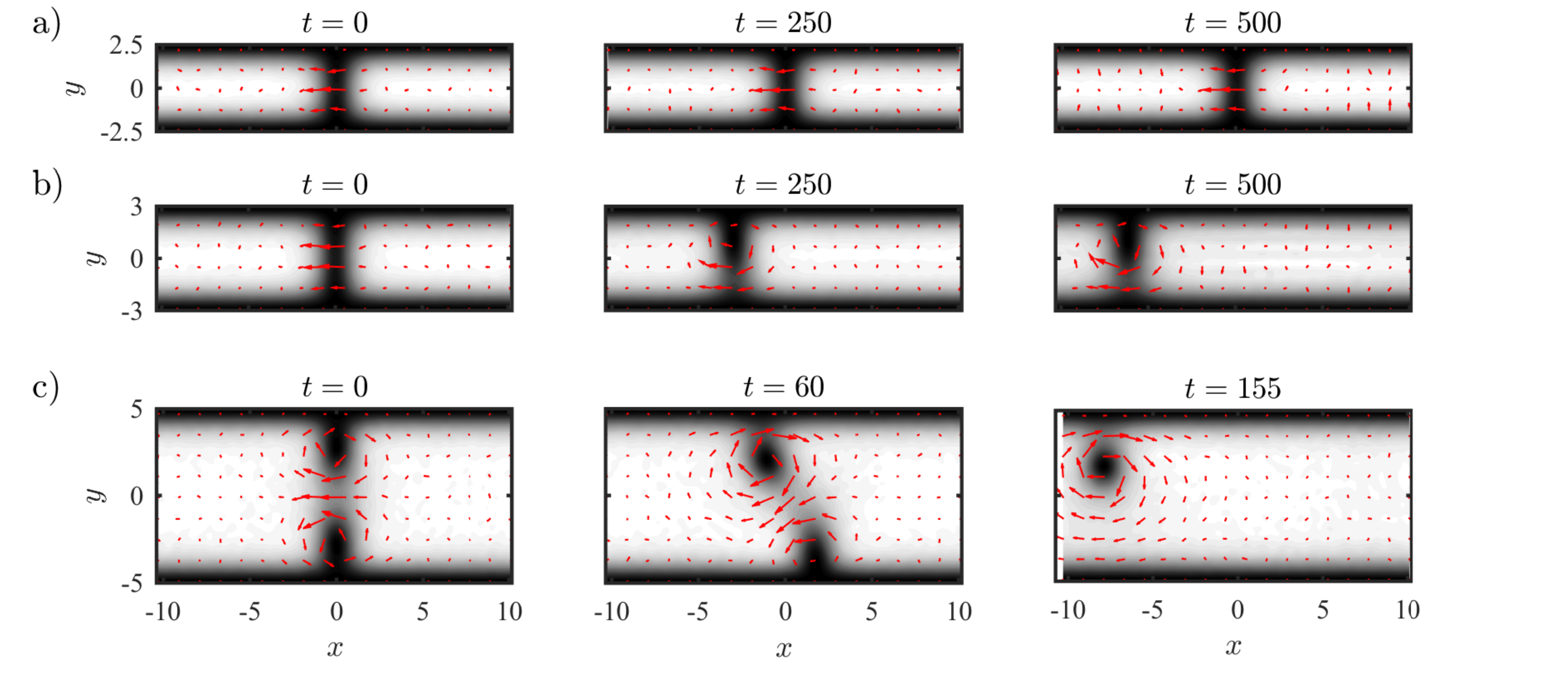}
  \caption{Stable and unstable dynamics for initially
    perturbed 1-lobed CDS solutions of speed $c =0.25$ evolved
    according to the NLS equation \eqref{eq:1}.  a) Stable dynamics
    for $L_y = 5$.  b) Unstable dynamics leading to a single vortex
    for $L_y = 6$.  c) 1-lobe CDS decay into two vortices resulting
    from the wider channel $L_y = 10$. The mean zero noise variance is
    $\sigma = 10^{-4}$.  Note that $\Lcr(0.25) \approx 5.68$ so the
    instability transition is accurately resolved.}
  \label{fig:dynamics_c_25}
\end{figure*}


\section{Conclusions\label{sec:conclusions}}%

The critical confinement widths for stabilizing propagating dark
solitons in the defocusing/repulsive NLS/GP equation were obtained
using spectral analysis and verified by direct computations.  The
results show that: (\emph{i}) for a given confinement width, the
faster the solitons propagate, the more stable they become;
(\emph{ii}) impenetrable walls are more stabilizing than zero flux
boundaries.  These results generalize upon previous studies for
black (non-propagating) dark solitons and gray (propagating) dark
solitons.  As part of this analysis, we also analytically obtained
approximate confined dark solitons with impenetrable walls.  This
approximation was used to show that confined dark solitons have a
reduced speed compared with the unconfined case, which is consistent
with and may help to explain experimental observations in BECs.

\appendix

\section*{Computational methods}

\section{Computing the CDS}
\label{sec:numer-comp}

We compute CDS bound state solutions of~\eqref{eq:u}
and~\eqref{eq:Dirichlet} using a spectrally accurate quasi-Newton
approach~\cite{Kelley-03}.  It is convenient to break up \eqref{eq:u}
into its real and imaginary parts by taking $\ucds = u + i v$ where
$u$, $v$ are real valued
\begin{equation}
  \label{eq:11}
  \begin{split}
    F(u,v) = \frac{1}{2} \Delta u + c v_\xi - (u^2 + v^2 - \mu) u &= 0, \\
    G(u,v) = \frac{1}{2} \Delta v - c u_\xi - (u^2 + v^2 - \mu) v &= 0 .
  \end{split}
\end{equation}
We seek solutions that rapidly asymptote to the uniform background
state 
\begin{equation}
  \label{eq:12}
  u^2(\xi,y) + v^2(\xi,y) \to u_\mathrm{b}(y)^2, \quad |\xi| \to
  \infty . 
\end{equation}
The far field density normalization \eqref{eq:normalize} determines the
chemical potential $\mu$ to be its background value \eqref{eq:7}.  

We solve eq.~\eqref{eq:11} using discrete cosine/sine transforms
achieving rapid (spectral) convergence.  For completeness, we include
a discussion of these transforms in the following subsection.  

\subsection{Discrete transforms}
\label{sec:comp-impl}

In order to evaluate eq.~(\ref{eq:u}) on a finite grid with spectral
accuracy, we introduce the half grid points on
$[-L_{\xi,y}/2,L_{\xi,y}/2]$
\begin{equation}
  \label{eq:6}
  \begin{split}
    \xi_i &= -\frac{L_{\xi}}{2} + \frac{2i - 1}{2} \Delta_\xi, \quad \Delta_\xi =
    \frac{L_{\xi}}{N_{\xi}}, \quad i = 1, 2, \ldots, N_{\xi} , \\
    y_j &= -\frac{L_y}{2} + \frac{2j - 1}{2} \Delta_y, \quad \Delta_y =
    \frac{L_y}{N_y}, \quad j = 1, 2, \ldots, N_y . 
  \end{split}
\end{equation}
The truncation of $\R$ to $\xi \in (-\Lxih,\Lxih)$ is achieved by
employing Neumann boundary conditions
\begin{equation}
  \label{eq:7}
  \begin{split}
    u_\xi(\pm \Lxih,y) = 0, \quad v_\xi(\pm \Lxih,y) = 0 ,
  \end{split}
\end{equation}
which are natural for dark solitons that rapidly decay to differing
constant values for $x \to \pm \infty$.  To evaluate derivatives and
simultaneously satisfy the boundary conditions, we approximate the
solution with a truncated cosine series expansion in $\xi$ and a
truncated sine series expansion in $y$
\begin{equation}
  \label{eq:8}
  \begin{split}
    &u(\xi, y) \approx \\
    &\frac{2}{(N_\xi N_y)^{1/2}}
    \sideset{}{'}{\sum}_{n=1}^{N_x} \sum_{m=1}^{N_y} \hat{u}_{n,m}
    \cos \left ( \frac{\pi(\xi + L_\xi/2)(n-1)}{L_\xi} \right ) \\
    &\qquad \qquad \qquad \qquad \qquad \times \sin
    \left ( \frac{\pi(y +\Lyh)(m-1)}{L_y}\right ) , \\
    &v(\xi, y) \approx \\
    &\frac{2}{(N_\xi N_y)^{1/2}}
    \sideset{}{'}{\sum}_{n=1}^{N_x} \sum_{m=1}^{N_y} \hat{v}_{n,m}
    \cos \left ( \frac{\pi(\xi + L_\xi/2)(n-1)}{L_\xi} \right ) \\
        &\qquad \qquad \qquad \qquad \qquad \times \sin
    \left ( \frac{\pi(y +\Lyh)(m-1)}{L_y}\right ) .
  \end{split}
\end{equation}
The prime on the first summation implies that whenever $n=1$, the
coefficient should be divided by $\sqrt{2}$.  The Fourier coefficients
are computed using fast cosine and sine transforms (DCT-II and DST-II)
\cite{Wang-AMC-85} from the discrete function values at the
half grid points.  The fast transforms are implemented by appropriate
pre and post processing of the FFT.

Derivatives are approximated according to
\begin{align}
  \label{eq:9a}
  u_{\xi} &\approx \mathcal{S}_{k_\xi}^{-1} \left \{ -k_\xi
    \mathcal{C}_\xi \left \{ \ucds
     \right \} \right \}, \\
  u_{\xi\xi} &\approx \mathcal{C}_{k_\xi}^{-1} \left \{ -k_\xi^2
    \mathcal{C}_\xi \left \{ \ucds
    \right \} \right \}, \\
  u_{yy} &\approx \mathcal{S}_{k_y}^{-1} \left \{ -k_y^2 \mathcal{S}_y
    \left \{ \ucds \right \} \right \} ,
\end{align}
where the notation $\mathcal{C}_\xi$ refers to a DCT in $\xi$ and
$\mathcal{S}_{k_\xi}^{-1}$ refers to an inverse DST in $k_\xi$, etc.
The discrete wavenumbers are
\begin{equation}
  \label{eq:10a}
  k_{\xi,y} \in \frac{\pi}{L_{\xi,y}} \{ 0, 1, \ldots, N_{\xi,y}\} .
\end{equation}

We use a black-box Newton solver on a preconditioned version of
eq.~\eqref{eq:11}, seeking $u$, $v$ such that $L^{-1} \mathbf{F}(u,v)
= 0$.  The preconditioner we use is
\begin{equation}
  \label{eq:11a}
  \begin{split}
    L &=
    \begin{bmatrix}
      a - \mu - \frac{1}{2} \Delta & - c \partial_{\xi} \\
      c \partial_{\xi} & a - \mu - \frac{1}{2} \Delta
    \end{bmatrix}, \\
    L^{-1} &= 
    \left [\left (a - \mu - \frac{1}{2} \Delta \right )^2 +
      c^2 \partial_{\xi\xi} \right ]^{-1}  \\
    & \quad\times
    \begin{bmatrix}
      a - \mu - \frac{1}{2} \Delta & - c \partial_{\xi} \\
      c \partial_{\xi} & a - \mu - \frac{1}{2} \Delta
    \end{bmatrix} ,
  \end{split}
\end{equation}
where $a > 0$ is an acceleration parameter.  The preconditioner is
applied efficiently using the DCST
\begin{widetext}
  \begin{equation}
  \label{eq:12a}
  L^{-1} \mathbf{F} =
  \begin{bmatrix}
    \mathcal{S}_{k_y}^{-1} \mathcal{C}_{k_\xi}^{-1} \left \{ \frac{a -
        \mu + \frac{1}{2}
        (k_\xi^2 + k_y^2) }{\left ( a - \mu +\frac{1}{2} (k_\xi^2 +
          k_y^2) \right )^2 - c^2 k_\xi^2} \mathcal{S}_{y}
      \mathcal{C}_{\xi} \left \{
        F \right \} \right \} + \mathcal{S}_{k_y}^{-1}
    \mathcal{S}_{k_\xi}^{-1} \left \{ 
      \frac{-c k_\xi }{\left ( a - \mu +\frac{1}{2} (k_\xi^2 +
          k_y^2) \right )^2 - c^2 k_\xi^2} \mathcal{S}_{y}
      \mathcal{C}_{\xi} \left \{ 
        G \right \} \right \} \\
    \mathcal{S}_{k_y}^{-1} \mathcal{C}_{k_\xi}^{-1} \left \{ \frac{a -
        \mu + \frac{1}{2}
        (k_\xi^2 + k_y^2) }{\left ( a - \mu +\frac{1}{2} (k_\xi^2 +
          k_y^2) \right )^2 - c^2 k_\xi^2} \mathcal{S}_{y}
      \mathcal{C}_{\xi} \left \{
        G \right \} \right \} - \mathcal{S}_{k_y}^{-1}
    \mathcal{S}_{k_\xi}^{-1} \left \{ 
      \frac{-c k_\xi }{\left ( a - \mu +\frac{1}{2} (k_\xi^2 +
          k_y^2) \right )^2 - c^2 k_\xi^2} \mathcal{S}_{y}
      \mathcal{C}_{\xi} \left \{ 
        F \right \} \right \}
  \end{bmatrix} .
\end{equation}
\end{widetext}

\begin{figure}
  \centering
  \includegraphics[scale=0.2]{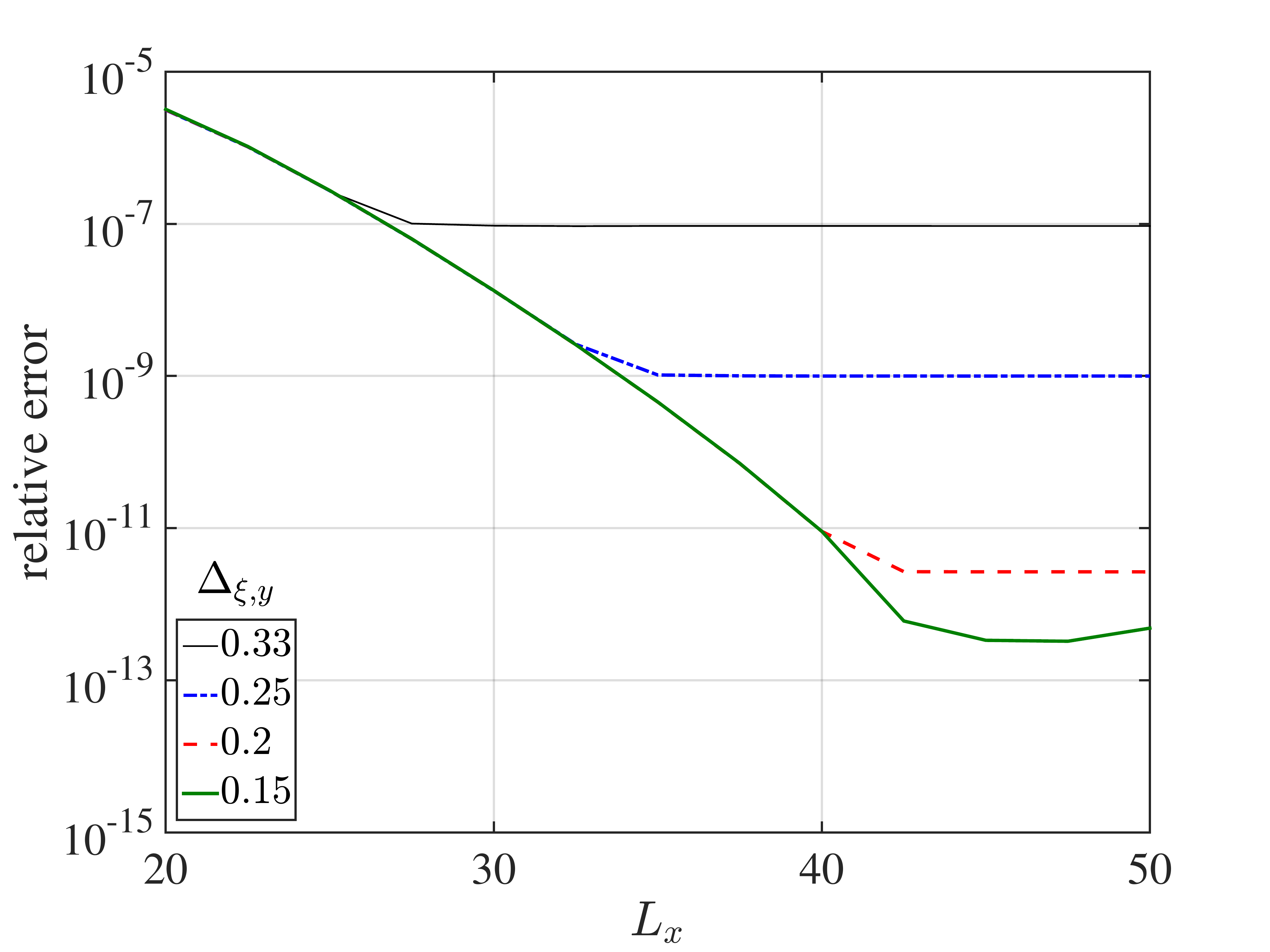}
  \caption{(Color online) Typical convergence of Newton calculations
    with $c = 0.5$, $L_y = 7$ with varying grid spacing
    ($\Delta_{\xi,y}$ in the legend) and longitudinal length $L_x$.  }
  \label{fig:newton_convergence}
\end{figure}

The stopping (convergence) criteria for Newton's method is an $l^2$
norm of the residual less than $10^{-13}$.  An example convergence
plot is presented in Fig.~\ref{fig:newton_convergence}.  The reference
``exact'' solution is computed on a large ($L_x = 60$), very fine grid
($\Delta = 0.1$).  The black soliton ($c = 0$) exhibits the strongest
localization and requires $L_x \approx 40$ to achieve the highest
accuracy.  The gray soliton with $c = 0.5$ is broader than the dark
soliton and hence requires a slightly longer channel width $L_x
\approx 50$ to achieve the highest accuracy.  For both the black and
gray solitons a grid spacing of $\Delta x = \Delta y = 0.15$ achieves
an absolute accuracy of $10^{-13}$.  For all calculations in this
study, we used $L_x = 40$ and grid spacing $\Delta = 0.2$, ensuring
absolute errors below $10^{-10}$.



\section{Linearization}
\label{sec:linearization}

Given a stationary solution of (\ref{eq:1}) via eq.~(\ref{eq:u}), its
linearization is of interest for Newton solver implementations and for
computing spectral stability.  Inserting the ansatz
\begin{equation}
  \label{eq:34}
  \begin{split}
    \psi(x,y,t) &= \big [ u(\xi,y) + \eps f(\xi,y) e^{\Gamma t} \\
    &\qquad + i
      \left ( v(\xi,y) + \eps g(\xi,y) e^{\Gamma t} \right ) \big
    ] e^{- i \mu t},
  \end{split}
\end{equation}
into eq.~(\ref{eq:1}) with $u$, $v$ a solution of \eqref{eq:11}
$F(u,v) = G(u,v) = 0$ and keeping only $\mathcal{O}(\eps)$ terms
results in the linearization
\begin{equation}
  \label{eq:35}
  \Gamma
  \begin{bmatrix}
    f \\ g
  \end{bmatrix}
  =
  \begin{bmatrix}
    0 &  - 1 \\
    1 & 0
  \end{bmatrix}
  \mathcal{J}(u,v)
  \begin{bmatrix}
    f \\ g
  \end{bmatrix},
\end{equation}
where $\mathcal{J} = \partial [F,G]/\partial [u,v]$ is the Jacobian of
the nonlinear system (\ref{eq:u})
\begin{equation}
  \label{eq:36}
  \mathcal{J}(u,v) =
  \begin{bmatrix}
    \frac{1}{2} \Delta - (3u^2 + v^2 - \mu) & c \partial_\xi - 2u
    v \\
    -c \partial_\xi - 2 u v & \frac{1}{2} \Delta - (u^2 + 3v^2
    - \mu)
  \end{bmatrix} .
\end{equation}
Note that $\mathcal{J}$ is self-adjoint.

\section{Dynamical evolution}
\label{sec:dynamical-evolution}

In this section we present a numerical scheme to efficiently solve
the NLS/GP equation~\eqref{eq:1} for the case of a channel 
with impenetrable walls~\eqref{eq:Dirichlet}.  
We use a split-step pseudospectral method.

We consider~\eqref{eq:1} in the reference frame moving with the
background flow of velocity $c$ in the $x$ direction
\begin{equation}
  \label{eq:24}
  \begin{split}
    i \psi_t &= -\frac{1}{2} (\psi_{xx} + \psi_{yy}) + i c \psi_x + |
    \psi |^2 \psi, \\
    \psi(x,y,0) &= \psi_0(x,y), \quad \psi(x,\pm\Lyh,t) = 0 .
  \end{split}
\end{equation}
The far field behavior \eqref{eq:ub1} implies non-homogeneous Dirichlet
boundary conditions for $\psi$.  However, for splitting methods, it is typically
advantageous to use homogeneous boundary conditions.  To this end, we
introduce Neumann boundary conditions in the $x$ direction as
\begin{equation}
  \label{eq:27}
  \psi_x(\pm \Lxh,y,t) = 0~.
\end{equation}
If we assume sufficient localization in the $x$ direction, the
condition \eqref{eq:27} is consistent with the background
state~\ref{eq:ub1} for sufficiently large $L_x$.

The standard split-step method is comprised of two steps as follows.
\begin{equation}
  \label{eq:43}
  \textrm{Nonlinear step:} \quad i \psi_t = |\psi|^2 \psi , \quad
  \psi(x,y,0) = \psi_0(x,y)
\end{equation}
and
\begin{equation}
  \label{eq:23}
  \begin{split}
    &\textrm{Linear step:} \quad
    i \psi_t = -\frac{1}{2} (\psi_{xx} + \psi_{yy}) + i c \psi_x, \\
    &\psi(x,y,0) = \psi_0(x,y), \quad \psi(x,\pm\Lyh,t) = 0, \\
    &\psi_x(\pm \Lxh,y,t) = 0 ,
  \end{split}
\end{equation}
The nonlinear step is an ODE, which can be solved exactly as
\begin{equation}
  \label{eq:46}
  \psi(x,y,t) = e^{-i |\psi_0(x,y)|^2 t} \psi_0(x,y).
\end{equation}
The linear step can be solved with spectral accuracy as 
\begin{align}
\nonumber
  \psi(x,y,t) &= e^{i t \Delta/2} \psi_0(x+ct,y) \\
  &= \mathcal{C}_{k_x}^{-1} \mathcal{S}_{k_y}^{-1} \Bigg \{ e^{- i
    t(k_x^2 + k_y^2)/2} \mathcal{C}_x \mathcal{S}_y \bigg \{
  \psi_0(\cdot+ct,\cdot) \bigg \}  \Bigg \} .
  \label{eq:47}
\end{align}
where trigonometric interpolation is used in order to evaluate
$\psi_0(x+ct,y)$.  The solutions of these two problems are then
chained together in order to achieve $2^{\mathrm{nd}}$ order accuracy
in time by taking successive half nonlinear, full linear, and half nonlinear steps.


%

\end{document}